# Discovery of Four Pulsars in a Pilot Survey at Intermediate Galactic Latitudes with FAST

Q. J. Zhi[1], J. T. Bai[2], S. Dai[3], X. Xu[4], S. J. Dang[1], L. H. Shang[1], R. S. Zhao[1], D. Li[5,6,7], W. W. Zhu[5], N. Wang[2,8,9], J. P. Yuan[2,8,9], P. Wang[5], L. Zhang[5], Y. Feng[10], J. B. Wang[11], S. Q. Wang[2,8,9], Q. D. Wu[2,12], A. J. Dong[1], H. Yang[1], J. Tian[1], W. Q. Zhong[1], X. H. Luo[1], Miroslav D. Filipović[3], and G. J. Qiao[13]

[1] School of Physics and Electronic Science, Guizhou Provincial Key Laboratory of Radio Astronomy and Data Processing, Guizhou Normal University, Guiyang 550001, People's Republic of China; qjzhi@gznu.edu.cn
[2] Xinjiang Astronomical Observatory, Chinese Academy of Sciences, Urumqi, Xinjiang 830011, People's Republic of China
[3] School of Science, Western Sydney University, Locked Bag 1797, Penrith South DC, NSW 2751, Australia
[4] School of Mathematical Sciences, Guizhou Provincial Key Laboratory of Radio Astronomy and Data Processing, Guizhou Normal University, Guiyang 550001, People's Republic of China
[5] National Astronomical Observatories, Chinese Academy of Sciences, Beijing 100101, People's Republic of China
[6] University of Chinese Academy of Sciences, Beijing 100049, People's Republic of China
[7] Research Center for Intelligent Computing Platforms, Zhejiang Laboratory, Hangzhou 311100, People's Republic of China
[8] Key Laboratory of Radio Astronomy, Chinese Academy of Sciences, Urumqi, Xinjiang 830011, People's Republic of China
[9] Xinjiang Key Laboratory of Radio Astrophysics, 150 Science1-Street, Urumqi, Xinjiang 830011, People's Republic of China
[10] Zhejiang Lab, Hangzhou, Zhejiang 311121, People's Republic of China
[11] Institute of Optoelectronic Technology, Lishui University, Lishui 323000, People's Republic of China
[12] School of Astronomy and Space Science, University of Chinese Academy of Sciences, Beijing 100049, People's Republic of China
[13] School of Physics, Peking University, Beijing 100871, People's Republic of China
*Received 2023 September 6; revised 2023 November 8; accepted 2023 November 9; published 2023 December 28*

## Abstract

We present the discovery and timing results of four pulsars discovered in a pilot survey at intermediate Galactic latitudes with the Five-hundred Aperture Spherical Telescope (FAST). Among these pulsars, two belong to the category of millisecond pulsars (MSPs) with spin periods of less than 20 ms. The other two fall under the classification of "mildly recycled" pulsars, with massive white dwarfs as companions. Remarkably, this small survey, covering an area of 4.7 deg$^2$, led to the discovery of four recycled pulsars. Such success underscores the immense potential of future surveys at intermediate Galactic latitudes. In order to assess the potential yield of MSPs, we conducted population simulations and found that both FAST and Parkes new phased array feed surveys, focusing on intermediate Galactic latitudes, have the capacity to uncover several hundred new MSPs.

*Unified Astronomy Thesaurus concepts:* Radio pulsars (1353)

## 1. Introduction

Radio pulsars, which are rapidly rotating neutron stars (NSs) emitting radio pulses, provide an exceptional opportunity for investigating profound aspects of physics. These include exploring gravitational theories in intensely strong fields (e.g., Venkatraman Krishnan et al. 2020; Kramer et al. 2021) and gaining insights into the state of matter under incredibly high densities (e.g., Demorest et al. 2010; Antoniadis et al. 2013; Özel & Freire 2016; Fonseca et al. 2021). Pulsars also enable us to investigate the characteristics of the interstellar medium (e.g., Coles et al. 2015; Abbate et al. 2020; Kumamoto et al. 2021; Zhang et al. 2023) and examine the formation and evolution of NSs within binary and other dense systems (e.g., Bhattacharya & van den Heuvel 1991). Among these pulsars, a particularly crucial group is the millisecond pulsars (MSPs), with spin periods usually shorter than approximately 20 ms. The remarkable stability of their rotation allows for exceptional timing precision, enabling us to detect ultra-low-frequency gravitational waves (e.g., Agazie et al. 2023; Antoniadis et al. 2023; Reardon et al. 2023; Xu et al. 2023). The extensive applications of pulsars in astrophysics have made the search for new pulsars a fundamental focus for current and future large radio telescopes (Nan 2006; Keane et al. 2015; Padmanabh et al. 2023; Wang et al. 2023).

While the majority of pulsars are situated near the Galactic plane, it is widely recognized that MSPs, or more generally "recycled" pulsars, exhibit significantly greater scale heights than regular pulsars (e.g., Levin et al. 2013). As we venture away from the Galactic plane, the density of the interstellar medium decreases, leading to reduced effects like scattering and dispersive smearing. This significant decrease in these effects enhances our ability to detect rapidly spinning MSPs, making our observations much more sensitive. Consequently, researchers have identified intermediate Galactic latitudes as the optimal region for MSP searches (Levin et al. 2013; McEwen et al. 2020). In the past, the Parkes High Time Resolution Universe (HTRU) survey (Keith et al. 2010) and drift-scan surveys conducted by various telescopes have covered these intermediate latitudes (e.g., Boyles et al. 2013; Deneva et al. 2013). Thanks to these surveys, the discovery of 79 MSPs has been achieved within $5° < |gb| < 15°$ (excluding globular clusters; see the ATNF pulsar catalog,[14] Manchester et al. 2005).

Currently, the Five-hundred-meter Aperture Spherical Telescope (FAST) is carrying out two major surveys with a focus on radio pulsars, the Commensal Radio Astronomy Fast Survey (CRAFTS; Li et al. 2018) and the Galactic Plane Pulsar Survey (GPPS; Han et al. 2021). So far, CRAFTS has discovered 179 radio pulsars, including 45 MSPs[15] (Qian et al. 2019; Zhang et al. 2019; Cameron et al. 2020; Cruces et al. 2021; Wang et al. 2021a, 2021b; Tedila et al. 2022; Wen et al. 2022; Miao et al. 2023; Wu et al. 2023), and GPPS has discovered 618 radio pulsars, including 148 MSPs[16] (Han et al. 2021; Zhou et al. 2023; Su et al. 2023). While these two surveys complement each other and GPPS is expected to cover Galactic latitudes up to 10°, a large fraction of intermediate Galactic latitudes (10°–15°) that are particularly rich in good-timer MSPs are still not covered with sufficient sensitivity. In this paper, we present the discovery of four pulsars in a pilot survey at intermediate Galactic latitudes and the results from our initial follow-up observations. In Section 2 we describe the survey and the timing campaign. In Section 3 we present the results and discuss their implications. Some perspectives are discussed in Section 4.

---
[14] https://www.atnf.csiro.au/research/pulsar/psrcat/

## 2. Observations and Data Reduction

### 2.1. A Pilot Survey at Intermediate Galactic Latitudes with FAST

The 19-beam $L$-band focal plane array of FAST (Li et al. 2018) was used to survey an area along the Galactic plane at a Galactic latitude of $gb = 5°.2$. The observing band covers a frequency range from 1.05 to 1.45 GHz (Jiang et al. 2020). A total of 30 pointings were carried out, and we listed their pointing centers in Table 1. We utilized the FAST snapshot observing mode (Han et al. 2021) for the survey, and therefore each pointing consists of four observations offset from each other to fully cover the region. The integration time of each observation is 390 s. The full pilot survey consists of 2280 beams. The FAST ROACH backend was used in its pulsar-search mode, with 4096 channels across 500 MHz of bandwidth and a 49 $\mu$s sampling rate. The total intensity was recorded with 8-bit sampling.

A periodicity search was carried out with the pulsar searching software package PRESTO (Ransom 2001). The dispersion measure (DM) range that we searched was 0–1000 pc cm$^{-3}$. In order to account for possible orbital modulation of pulsar periodic signals, we searched for signals drifting by as much as $\pm 200/n_h$ bins in the Fourier domain by setting zmax = 200 (Ransom et al. 2002), where $n_h$ is the largest harmonic at which a signal is detected (up to eight harmonics were summed). We also searched for single-pulse candidates with a signal-to-noise ratio (S/N) larger than seven using the single_pulse_search.py routine for each dedispersed time series and boxcar filtering parameters with filter widths ranging from 1 to 300 samples. Burst candidates were manually examined, and narrowband and impulsive radiofrequency interference were manually removed.

Four pulsar candidates were detected with our periodicity search. Follow-up observations of these candidates were performed using FAST and the Parkes radio telescope, Murriyang. All four candidates were successfully confirmed. In Table 2, we present the measured parameters of these pulsars.

---
[15] http://groups.bao.ac.cn/ism/english/CRAFTS/202210/t20221026_719407.html
[16] http://zmtt.bao.ac.cn/GPPS/GPPSnewPSR.html

**Table 1**
A List of Covers by the Survey at Intermediate Galactic Latitudes

| Pointing | GL (deg) | GB (deg) | R.A. (J2000) (hh:mm:ss) | Decl. (J2000) ($\pm$ dd:mm:ss) |
|---|---|---|---|---|
| 1 | 26.215 | 5.233 | 18:20:30.47 | −03:33:31.4 |
| 2 | 27.025 | 5.233 | 18:22:00.38 | −02:50:38.6 |
| 3 | 27.835 | 5.233 | 18:23:29.97 | −02:07:43.7 |
| 4 | 28.645 | 5.233 | 18:24:59.24 | −01:24:46.8 |
| 5 | 29.455 | 5.233 | 18:26:28.24 | −00:41:47.9 |
| 6 | 30.265 | 5.233 | 18:27:57.00 | +00:01:12.7 |
| 7 | 31.075 | 5.233 | 18:29:25.23 | +00:44:15.1 |
| 8 | 31.885 | 5.233 | 18:30:53.86 | +01:27:19.2 |
| 9 | 32.695 | 5.233 | 18:32:22.04 | +02:10:24.7 |
| 10 | 33.505 | 5.233 | 18:33:50.07 | +02:53:31.6 |
| 11 | 34.315 | 5.233 | 18:35:17.99 | +03:36:39.8 |
| 12 | 35.125 | 5.233 | 18:36:45.83 | +04:19:49.1 |
| 13 | 35.935 | 5.233 | 18:38:13.61 | +05:02:59.5 |
| 14 | 36.745 | 5.233 | 18:39:41.36 | +05:46:10.9 |
| 15 | 37.555 | 5.233 | 18:41:09.11 | +06:29:23.1 |
| 16 | 38.365 | 5.233 | 18:42:36.89 | +07:12:36.0 |
| 17 | 39.175 | 5.233 | 18:44:04.72 | +07:55:49.6 |
| 18 | 39.985 | 5.233 | 18:45:32.64 | +08:39:03.7 |
| 19 | 40.795 | 5.233 | 18:47:00.67 | +09:22:18.2 |
| 20 | 41.605 | 5.233 | 18:48:28.83 | +10:05:33.1 |
| 21 | 42.415 | 5.233 | 18:49:57.17 | +10:48:48.2 |
| 22 | 43.225 | 5.233 | 18:51:25.70 | +11:32:03.4 |
| 23 | 44.035 | 5.233 | 18:52:54.46 | +12:15:18.6 |
| 24 | 44.845 | 5.233 | 18:54:23.48 | +12:58:33.6 |
| 25 | 45.645 | 5.233 | 18:55:52.79 | +13:41:48.5 |
| 26 | 46.465 | 5.233 | 18:57:22.41 | +14:25:03.0 |
| 27 | 47.275 | 5.233 | 18:58:52.39 | +15:08:17.1 |
| 28 | 48.085 | 5.233 | 19:00:22.75 | +15:51:30.7 |
| 29 | 48.895 | 5.233 | 19:01:53.52 | +16:34:43.6 |
| 30 | 49.705 | 5.233 | 19:03:24.74 | +17:17:55.7 |

### 2.2. Follow-up Timing Observations

Two of the bright pulsars, J1826−0049 and J1849+1001, were followed up with the Parkes telescope. The ultra-wideband low system (Hobbs et al. 2020) was used in the coherently dedispersed search mode where data were recorded with 2-bit sampling every 64 $\mu$s in each of the 1 MHz wide frequency channels covering a total bandwidth of 3328 MHz between 704 and 4032 MHz. Only the total intensity was recorded. The integration time is 1 hr for J1826−0049 and 2 hr for J1849+1001. PSR J1839+0543 was observed and timed with FAST using the central beam of the 19-beam receiver. Data were recorded in the pulsar search mode with configurations the same as in the survey. Full polarization information was recorded. The integration time for each pulsar is 240 s.

To derive coherent timing solutions, search-mode data were folded with the apparent spin period of each pulsar determined at each observing epoch using the DSPSR software package (van Straten & Bailes 2011) with a subintegration length of 30 s. We manually excised data affected by narrowband and impulsive radiofrequency interference for each subintegration. Each observation was averaged in time to create subintegrations with a length of a few minutes, and pulse times of arrival (ToAs) were measured for each subintegration using the pat routine of the PSRCHIVE software package (van Straten et al. 2012). Timing analysis was carried out using the TEMPO2 software package (Hobbs et al. 2006). We used the barycentric coordinate time units, TT(TAI) clock standard, and JPL DE438 solar system ephemeris for our timing analysis.

**Table 2**
Parameters of Five Pulsars

| Pulsars with timing solutions | | |
|---|---|---|
| | J1826−0049 | J1849+1001 |
| RAJ (J2000) | 18:26:16.546(1) | 18:49:00.7303(4) |
| DECJ (J2000) | −00:49:50.07(5) | +10:01:01.07(2) |
| $\nu$ (Hz) | 217.8248734478(2) | 28.41772282880(1) |
| $\dot{\nu}$ (Hz s$^{-1}$) | $-1.1(2) \times 10^{-15}$ | $-4(1) \times 10^{-17}$ |
| PMRA (mas yr$^{-1}$) | 149(43) | |
| PMDEC (mas yr$^{-1}$) | 312(102) | |
| EPOCH (MJD) | 59971.98 | 59950.05 |
| Time span (MJD) | 59754−60189 | 59749−60133 |
| DM (cm$^{-3}$ pc) | 42.674(8) | 79.56(1) |
| Reduced $\chi^2$ | 1.5628 | 0.8893 |
| Wrms ($\mu$s) | 4.7 | 19.4 |
| Binary parameters | | |
| | ELL1 model | DD model |
| $P_b$ (days) | 6.73497259(4) | 26.1656358(1) |
| $\chi$ (ls) | 5.978283(2) | 44.751772(7) |
| $T_{ASC}$ (MJD) | 59972.1346047(5) | |
| $T_0$ (MJD) | | 59932.3275(1) |
| EPS1 | $-1.0(8) \times 10^{-6}$ | |
| EPS2 | $1.1(7) \times 10^{-6}$ | |
| OM | | 336.079(1) |
| ECC | | 0.0080940(2) |
| Derived parameters | | |
| GL (deg) | 29.313 | 41.596 |
| GB (deg) | 5.215 | 5.082 |
| $\dot{E}$ (erg s$^{-1}$) | $9.5 \times 10^{33}$ | $4.5 \times 10^{31}$ |
| $B_s$ (G) | $3.3 \times 10^8$ | $1.3 \times 10^9$ |
| $\tau_c$ (Myr) | 3000 | 11500 |
| DIST$_{YMW16}$ (kpc) | 1.3 | 3.2 |
| $P$ (ms) | 4.59084393858(5) | 35.18930795490(2) |
| $\dot{P}$ (s s$^{-1}$) | $2.3(4) \times 10^{-20}$ | $5(1) \times 10^{-20}$ |
| Companion mass ($M_\odot$) | 0.2332 <0.2738 <0.6185 | 0.8902 <1.0873 <3.3440 |
| Pulsars without timing solutions | | |
| | J1839+0543 | J1852+1200 | J1828−0003 |
| RAJ (J2000) | 18:39:00(20) | 18:52:58(20) | 18:28:44(20) |
| DECJ (J2000) | +05:43(2) | +12:00(2) | −00:03(2) |
| $\nu$ (Hz) | 17.26282301(1) | 258.687554(7) | 0.262666(7) |
| EPOCH (MJD) | 58447.77 | 60127 | 59310 |
| Time span (MJD) | 58643−59792 | | |
| DM (cm$^{-3}$ pc) | 113.83(1) | 68.05(5) | 193(3) |
| RM (rad m$^{-2}$) | 211(20) | | |
| Binary parameters | | | |
| $P_b$ (days) | 28.517(1) | | |
| $\chi$ (ls) | 46.55(3) | | |
| $T_0$ (MJD) | 59821.97(6) | | |
| OM | 166(8) | | |
| ECC | 0.0039(5) | | |
| Derived parameters | | | |
| GL (deg) | 36.620 | 43.812 | 30.292 |
| GB (deg) | 5.362 | 5.106 | 5.027 |
| DIST$_{YMW16}$ (kpc) | 5.3 | 2.7 | 10.6 |
| $P$ (ms) | 57.927974(5) | 3.8657782(1) | 3807.1(1) |
| Companion mass ($M_\odot$) | 0.8662 <1.0568 <3.2190 | | |

**Note.** The minimum, median, and maximum companion masses for binary systems were estimated assuming a pulsar mass of 1.35 $M_\odot$ and inclination angles of 90°, 60°, and 26°, respectively.

For pulsars for which a coherent timing solution could be obtained, we refolded the search-mode data and averaged each observation in time and frequency to produce a high S/N pulse profile. ToAs were measured using these high S/N profiles, and we repeated our timing analysis to measure their spin, astrometric, and binary parameters. Throughout our timing analysis, TEMPO2 fitting with ToA errors (known as "MODE 1") was used and the weighted rms (Wrms) of timing residuals were reported in Figure 2 and Table 2. To refine our DM measurements, for each pulsar we selected a high S/N observation and divided the bandwidth into four frequency subbands. We then measured a ToA from each subband and fitted for the DM using TEMPO2. Our timing results will be presented in Section 3.

To perform polarimetric calibration for FAST observations, we conducted noise diode observations prior to each observation. A 100% linearly polarized diode signal with a period of 0.100663296 s was injected into the receiver system as the telescope points toward a sky region offset by 10′ from the target source (Jiang et al. 2020). The PAC routine of PSRCHIVE was used to calibrate the polarization of each observation. The Stokes parameters are in accordance with the astronomical conventions described by van Straten et al. (2010). Stokes $V$ is defined as $I_{LH} - I_{RH}$, using the IEEE definition for the sense of circular polarization. The linear polarization and the position angle of linear polarization were calculated following Dai et al. (2015). After the polarimetic calibration, we searched for the Faraday rotation measure (RM) for each pulsar using the RMFIT routine of PSRCHIVE. We carried out a brute-force search for peak linear polarization with RMFIT in an RM range of ±1000 rad m$^{-2}$.

## 3. Results

With 13.5 hr of observing time that covers 4.7 deg$^2$ of area, we discovered four new pulsars and detected all seven known pulsars in this region. Two of the new discoveries (PSRs J1826−0049 and J1852+1200) are MSPs with spin periods shorter than 20 ms. Three of them (PSRs J1826−0049, J1849+1001, and J1839+0543) are confirmed to be in binary systems. So far, coherent timing solutions have been obtained for two pulsars (J1826−0049 and J1849+1001). PSR J1828−0003 was previously discovered as a rotating radio transient (RRAT; Zhou et al. 2023), and we detected its periodic signals with a period of 3.8071 s. In the following sections, we will discuss each pulsar separately.

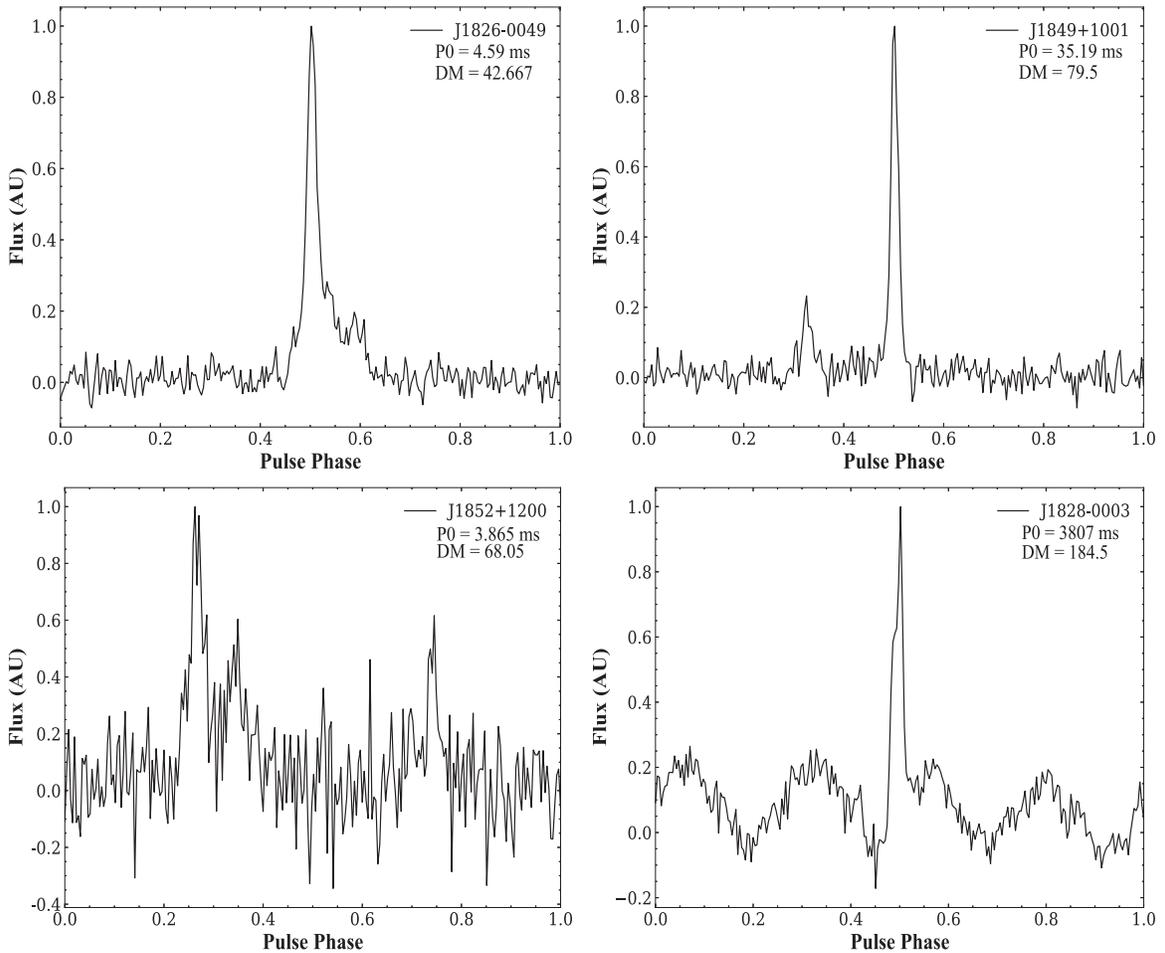

**Figure 1.** Time- and frequency-averaged pulse profiles of PSRs J1826−0049, J1849+1001, J1852+120, and J1828−0003.

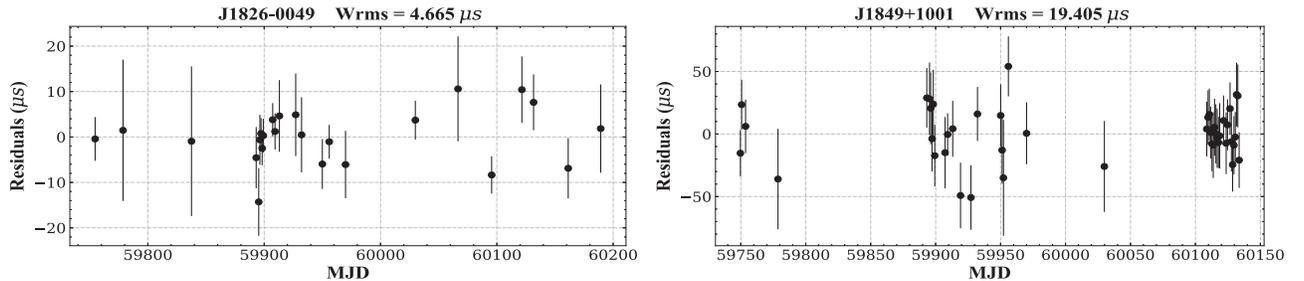

**Figure 2.** Timing residuals for two new pulsars as a function of MJD. The Wrms of the timing residuals of each pulsar is reported.

### 3.1. J1826−0049

PSR J1826−0049 is an MSP with a period of 4.59 ms and a DM of 42.67 pc cm$^{-3}$. In Figure 1 we show its time- and frequency-averaged pulse profile using a Parkes observation with 1 hr of integration. With follow-up Parkes timing observations, we successfully obtained a coherent timing solution for PSR J1826−0049 (Table 2). Best-fit timing residuals are shown in Figure 2. Our current timing showed that it is in a binary system with an orbital period of 6.7 days. The minimum, median, and maximum companion masses are 0.2332, 0.2738, and 0.6185 $M_\odot$, respectively. Here we assumed the pulsar mass to be $M_p = 1.35\ M_\odot$, and the minimum, median, and maximum companion masses were estimated with an inclination angle of $i = 90°$, $60°$, and $26°$, respectively (Lorimer & Kramer 2004). This suggests that the companion to PSR J1826−0049 is likely a white dwarf (WD). Our timing analysis yields a large proper motion of 346(94) mas yr$^{-1}$ for this system. However, for a DM distance of 1.3 kpc (Yao et al. 2017), this gives an apparent acceleration due to the proper motion (the so-called Shklovskii effect; Shklovskii 1970) of $\sim 1.1 \times 10^{-7}$ m s$^{-2}$, which corresponds to an apparent $\dot{P}_{\rm Shk}$ of $\sim 1.7 \times 10^{-18}$ s s$^{-1}$. This is almost two orders of magnitude larger than the measured $\dot{P}$ of $\sim 2.3 \times 10^{-20}$ s s$^{-1}$, suggesting that the proper motion was significantly overestimated. Continued timing observations and a longer timing baseline are required to constrain the proper motion of this pulsar.

### 3.2. J1849+1001

PSR J1849+1001 has a spin period of 35.189 ms and a DM of 79.56 pc cm$^{-3}$. In Figure 1 we show its time- and frequency-averaged pulse profile using a Parkes observation with 2 hr of integration. We carried out a high cadence timing campaign with Parkes in 2022 and successfully obtained a coherent timing solution for PSR J1849+1001 (Table 2). Best-fit timing residuals are shown in Figure 2. Our current timing showed that it is in a binary system with an orbital period of 26.16 days. The minimum, median, and maximum companion masses are 0.8902, 1.0873 and 3.3440 $M_\odot$, respectively. The system's low eccentricity (e ∼0.008) excludes the possibility that this is a double NS system; it is very likely to be a member of the subclass of mildly recycled pulsars and has a massive WD companion (e.g., Gautam et al. 2022).

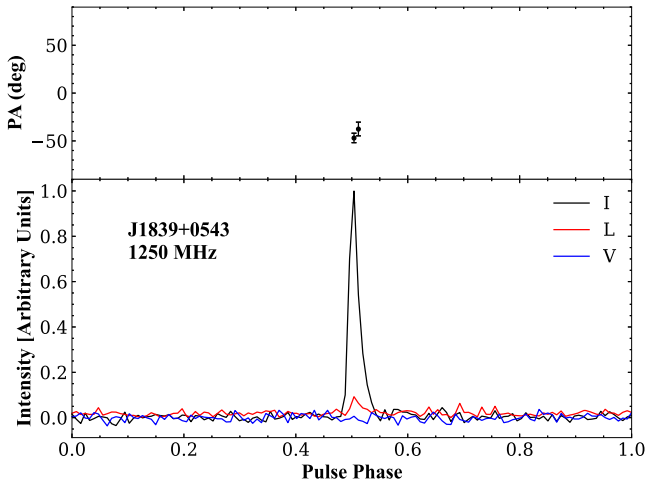

**Figure 3.** Polarization profile of PSR J1839+0543. The black, red, and blue lines show the total intensity, linear polarization, and circular polarization, respectively.

### 3.3. J1839+0543

PSR J1839+0543 has a spin period of 57.927 ms and a DM of 113.83 pc cm$^{-3}$. While a coherent timing solution has not been obtained so far, our follow-up FAST observations confirmed that it is in a binary system with an orbital period of 28.517 days. The minimum, median, and maximum companion mass are 0.87, 1.06 ,and 3.22 $M_\odot$, respectively. Similar to J1849+1001, the eccentricity of this system (e ∼0.004) is low and is likely to be a mildly recycled pulsar with a massive WD companion. Using measured pulsar parameters, we coadded all FAST observations and obtained a high S/N pulse profile, which enabled us to measure its RM to be 211 ± 20 rad cm$^{-2}$. In Figure 3 we show the time- and frequency-averaged polarization profile of PSR J1839+0543.

### 3.4. J1828−0003

PSR J1828−0003 was previously discovered as a RRAT by the FAST GPPS survey (Zhou et al. 2023). We detected nine single pulses (>9σ) at a DM of 193 pc cm$^{-3}$ from this pulsar through our single-pulse search. In Figure 4, we show our initial detection of single pulses of this pulsar. We searched for periodicity with detected single pulses and identified a period of 3.8 s. Subsequently, we folded our FAST data with the period and successfully detected its pulsed emission (shown in Figure 1). This confirmed that PSR J1828−0003 is likely to be an isolated normal pulsar with RRAT-like activities.

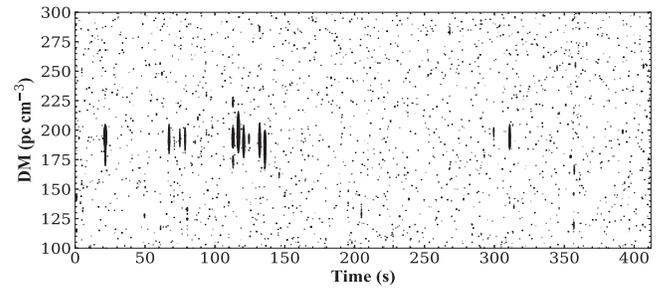

**Figure 4.** Detected single pulses of PSR J1828−0003. Each circle represents a detection above a threshold of 5σ. The diameter of each circle is proportional to the significance of the detection.

### 3.5. J1852+1200

PSR J1852+1200 is an MSP with a spin period of 3.866 ms and a DM of 68.05 pc cm$^{-3}$. While FAST observations have confirmed the discovery, timing observations of this pulsar have not yet commenced, leaving its binary system status unknown.

## 4. Discussion and Conclusion

It is generally believed that MSPs in binary systems have been "recycled" by the accretion of matter and transfer of angular momentum from their binary companion, spinning up their rotation to millisecond periods (e.g., Bhattacharya & van den Heuvel 1991). In some cases, this accretion process can stop before the pulsar gets fully recycled, leading to so-called mildly recycled pulsars with rotational periods between 20 and 100 ms. This process happens mostly if the companion stars are more massive: such stars evolve more rapidly, and therefore any accretion episodes will generally be much shorter (Berthereau et al. 2023). Two of our discoveries, PSRs J1849 +1001 and J1839+0543, fall under this category.

Their massive companions (>1 $M_\odot$) and low orbital eccentricities suggest that PSRs J1849+1001 and J1839 +0543 can be classified as intermediate-mass binary pulsars (see Tauris et al. 2012). More interestingly, our current analysis suggests that the WD companion of these two pulsars could be significantly more massive than 1 $M_\odot$. This means that PSRs J1849+1001 and J1839+0543 could be very similar to PSR J2045+3633 (McKee et al. 2020). Precise measurements of their companion masses and orbital parameters are therefore important for us to understand their evolutionary history and the general evolution of intermediate-mass binary pulsars.

Because of their massive companions and nonnegligible orbital eccentricity, PSRs J1849+1001 and J1839+0543 could be ideal systems to measure "post-Keplerian" parameters through pulsar timing (e.g., Damour & Taylor 1992). If we assume that general relativity adequately describes these effects, then two post-Keplerian parameters suffice to determine the masses of both components of a binary. Such measurements are of great importance for probing the equation of state of NSs. Currently, our timing baseline of approximately 1 yr is too short to measure any relativistic perturbations to the pulsar's orbit, and longer timing with high precision is required.

With subarcsec precision timing positions of PSRs J1826 −0049 and J1849+1001, we searched for their multiwavelength counterparts in publicly available optical, X-ray, and γ-ray surveys. No counterparts have been identified. Given their substantial Galactic latitudes, these binary systems represent promising candidates for dedicated deep optical observations. In addition, our current measurement of the $\dot{P}$ of J1826−0049 gives a large spin-down power of $\dot{E} \approx 9.5 \times 10^{33}$ erg s$^{-1}$, suggesting that this could also be an X-ray and/or γ-ray pulsar.

One of the motivations to find more MSPs is to improve the sensitivity of current pulsar timing arrays to detect ultra-low frequency gravitational waves (Xu et al. 2023; Reardon et al. 2023). PSR J1826−0049 is a comparatively bright MSP, which can be detected with high S/N by FAST in ∼5 minutes and by Parkes in ∼1 hr. Therefore, it can be a good candidate for pulsar timing arrays in the future. Our continued timing of this pulsar at Parkes will allow us to refine its parameters and evaluate its timing precision. The other MSP, J1852+1200, is much fainter than J1826−0049 and can only be timed by large telescopes like FAST.

The discovery of two MSPs (PSRs J1826−0049 and J1852 +1200) and two recycled pulsars (PSRs J1849+1001 and J1839+0543) in our pilot survey further stressed the importance of sensitive pulsar surveys at intermediate Galactic latitudes. To demonstrate this, we utilized the PSRPOPPY software package (Bates et al. 2014) to predict the number of MSP discoveries for FAST surveys covering $5° < |gb| < 10°$ and $10° < |gb| < 15°$, assuming that the integration time per pointing is identical to our pilot survey (i.e., 390 s). Here we followed the procedure described in Dai et al. (2017) to perform the population simulation and used MSP Galactic, spin period, luminosity, and spectral distributions presented by previous studies (e.g., Yusifov 2004; Faucher-Giguère & Kaspi 2006; Lorimer et al. 2006; Levin et al. 2013; Lorimer et al. 2015). Our simulations showed that ∼616 MSPs are expected to be detected by FAST within $5° < |gb| < 10°$ and ∼322 MSPs within $10° < |gb| < 15°$. So far, 25 and 12 Galactic MSPs have been discovered in these two regions, respectively.

A new cryogenically cooled phased array feed (cryoPAF) is currently being commissioned at the Parkes radio telescope. The cryoPAF has a field of view four times larger than the legacy Parkes multibeam receiver and therefore allows us to carry out much deeper pulsar surveys with the same amount of observing time as previous surveys. We carried out a simulation assuming that the cryoPAF repeats the Parkes HTRU mid-lat survey ($3°.5 < |gb| < 15°$) with an integration time of 2160 s, which is four times longer than that of HTRU mid-lat (Keith et al. 2010). Our simulation indicates that the cryoPAF survey is anticipated to identify approximately ∼160 MSPs. Currently, 85 Galactic MSPs have been detected in this specific area. Given the limited overlap in the sky coverage between FAST and Parkes, conducting a fresh survey at intermediate Galactic latitudes using the cryoPAF technology holds substantial promise and potential.

## Acknowledgments

This work made use of the data from FAST (Five-hundred-meter Aperture Spherical radio Telescope). FAST is a Chinese national megascience facility operated by the National Astronomical Observatories, Chinese Academy of Sciences. The Parkes radio telescope is part of the Australia Telescope National Facility, which is funded by the Commonwealth of Australia for operation as a national facility managed by CSIRO. This work is supported by the National Natural Science Foundation of China (grant Nos. 11988101, Z12273008, 12041303, 12041304), the National SKA Program of China (grant Nos. 2022SKA0130100, 2022SKA0130104, 2020SKA0120200), the Natural Science and Technology Foundation of Guizhou Province (grant No. [2023]024), the Foundation of Guizhou Provincial Education Department (grant No. KY (2020) 003), the Major Science and Technology Program of Xinjiang Uygur Autonomous Region (grant No. 2022A03013-3), the Academic New Seedling Fund Project of Guizhou Normal University (grant No. [2022]B18), and the Scientific Research Project of the Guizhou Provincial Education (grant Nos. KY[2022]137, KY[2022]132). S.D. is the recipient of an Australian Research Council Discovery Early Career Award (DE210101738) funded by the Australian Government. L.Z. is supported by ACAMAR Postdoctoral Fellowship and the National Natural Science Foundation of China (grant No. 12103069). Y.F. is supported by National Natural Science Foundation of China (grant No. 12203045).

## ORCID iDs

Q. J. Zhi https://orcid.org/0000-0001-9389-5197
J. T. Bai https://orcid.org/0000-0002-1052-1120
S. Dai https://orcid.org/0000-0002-9618-2499
X. Xu https://orcid.org/0009-0006-3224-4319
S. J. Dang https://orcid.org/0000-0002-2060-5539
L. H. Shang https://orcid.org/0000-0002-9173-4573
R. S. Zhao https://orcid.org/0000-0002-1243-0476
D. Li https://orcid.org/0000-0003-3010-7661
W. W. Zhu https://orcid.org/0000-0001-5105-4058
N. Wang https://orcid.org/0000-0002-9786-8548
J. P. Yuan https://orcid.org/0000-0002-5381-6498
P. Wang https://orcid.org/0000-0002-3386-7159
L. Zhang https://orcid.org/0000-0001-8539-4237
Y. Feng https://orcid.org/0000-0002-0475-7479
J. B. Wang https://orcid.org/0000-0001-9782-1603
S. Q. Wang https://orcid.org/0000-0003-4498-6070
A. J. Dong https://orcid.org/0000-0003-4151-0771
Miroslav D. Filipović https://orcid.org/0000-0002-4990-9288